\documentclass[a4paper,accepted=2025-11-03]{quantumarticle}
\pdfoutput=1
\usepackage[utf8]{inputenc}
\usepackage[english]{babel}
\usepackage[T1]{fontenc}	

\usepackage{graphicx}
\usepackage{amsmath}
\usepackage{amssymb}

\usepackage{hyperref}
\hypersetup{
 unicode=true,
 pdftoolbar=false, pdfmenubar=true, pdffitwindow=false, pdfstartview={FitH},
 pdftitle={Dicke subsystems are entangled},
 pdfauthor={Szilárd Szalay, Péter Nyári},
 pdfsubject={research article on multipartite entanglement in Dicke states},
 pdfkeywords={multipartite entanglement} {Dicke states},
 pdfcreator={pdflatex}, pdfproducer={overleaf},
 pdfnewwindow=true, colorlinks=true, linktoc=page, linkcolor=blue, citecolor=blue, filecolor=blue, urlcolor=blue
}

\usepackage[numbers,sort&compress]{natbib}

\DeclareMathOperator{\Tr}{Tr}

\newcommand{\cmpl}[1]{\overline{#1}}
\newcommand{\norm}[1]{\vert\vert #1 \vert\vert}
\newcommand{\abs}[1]{\vert #1 \vert}

\newcommand{\mi}[1]{{\boldsymbol{#1}}}

\newcommand{\ket}[1]{\vert #1 \rangle}
\newcommand{\bra}[1]{\langle #1 \vert}
\newcommand{\proj}[1]{\ket{#1}\bra{#1}}
\newcommand{\scalp}[2]{\langle #1 \vert #2 \rangle}

\newcommand{\set}[1]{\{ #1 \}}
\newcommand{\bigset}[1]{\bigl\{ #1 \bigr\}}

\newcommand{\bigsset}[2]{\bigset{ #1 \;\big\vert\; #2 }}

\newcommand{\DscGrp}[1]{\mathrm{#1}}

\newcommand{\T}{\mathrm{T}}
\newcommand{\D}{\text{D}}

\begin{document}

\title{Dicke subsystems are entangled}
\author{Szilárd Szalay}
\email{szalay.szilard@wigner.hu}
\affiliation{Department of Theoretical Solid State Physics,
HUN-REN Wigner Research Centre for Physics, Budapest, Hungary}
\affiliation{Department of Theoretical Physics,
University of the Basque Country UPV/EHU, Bilbao, Spain}
\affiliation{EHU Quantum Center, 
University of the Basque Country UPV/EHU, Leioa, Biscay, Spain}
\author{Péter Nyári}
\email{nyaripeter@student.elte.hu}
\affiliation{Department of Theoretical Solid State Physics,
HUN-REN Wigner Research Centre for Physics, Budapest, Hungary}
\affiliation{Department of Physics of Complex Systems,
Eötvös Loránd University, Budapest, Hungary}

\date{November 5, 2025}

\begin{abstract}
We show that all reduced states of nonproduct symmetric Dicke states of arbitrary number of qu$d$its
are genuinely multipartite entangled,
and of nonpositive partial transpose with respect to any subsystem.
\end{abstract}

\maketitle

Entanglement is the most remarkable manifestation of the nonclassical correlations in quantum systems~\cite{Werner-1989,Horodecki-2009},
still puzzling the community for almost hundred years~\cite{Schrodinger-1935a}.
Deciding if a mixed quantum state is entangled or separable is a notoriously difficult problem,
most of the known criteria are necessary but not sufficient for separability~\cite{Horodecki-2009,Guhne-2009,Szalay-2011}.
A particularly strong separability criterion is the Peres-Horodecki, or partial transpose criterion~\cite{Peres-1996,Horodecki-1996}.
It states that
if a quantum state is separable then its partial transposition is positive semidefinite (PPT),
or contrapositively,
if the partial transposition of a quantum state is not positive semidefinite (NPT) then the state is entangled.
Since the reverse implication is not true in general,
there are also entangled states of positive partial transpose~\cite{Horodecki-1997,Szalay-2011,Huber-2018,Pal-2019,Krebs-2024}.
This PPT-entanglement is considered to be a weaker form of entanglement.

A bipartite quantum state can either be separable or entangled~\cite{Werner-1989,Bengtsson-2006,Horodecki-2009},
while in the multipartite case entanglement shows a rich structure with many exciting features~\cite{Bengtsson-2017b,Horodecki-2024}.
A natural generalization of the separable/entangled dichotomy to multipartite systems
is the partial separability classification~\cite{Dur-1999,Acin-2001,Seevinck-2008,Szalay-2015b,Szalay-2019,Szalay-2025},
including the notions of \emph{partition-separability},
\emph{$k$-separability}, \emph{$k$-producibility}, \emph{$k$-stretchability}~\cite{Szalay-2019,Szalay-2025},
\emph{biseparability} or \emph{genuine multipartite entanglement} (GME).
Deciding if a mixed quantum state possesses a particular multipartite entanglement property is an even more difficult problem
than in the bipartite case.

A particularly interesting problem in the multipartite setting is that
the entanglement inside the subsystems does not follow from the entanglement of the whole system.
The paradigmatic example of this is given by the three-qubit GHZ state $\bigl(\ket{000}+\ket{111}\bigr)/\sqrt{2}$
and W state $\bigl(\ket{001}+\ket{010}+\ket{100}\bigr)/\sqrt{3}$,
both of which are fully entangled, 
but the two-qubit subsystems of the GHZ state are separable,
while those of the W state are entangled~\cite{Dur-2000b}.
Symmetric Dicke states are the generalizations of the W state,
and our result is the generalization of this.

In this work we consider mixed quantum states arising as reduced states of pure multiqu$d$it symmetric Dicke states,
and we show that these states are genuinely multipartite entangled,
moreover, this entanglement is also NPT (not PPT) with respect to any subsystem.
This holds for all nontrivial cases, i.e.,
when there are at least two nonzero occupations in the Dicke state of the whole system.

Dicke states originally appeared in quantum optics~\cite{Dicke-1954},
and later, thanks to their simple yet interesting structure,
became widely used examples and tools in the theory of multipartite entanglement.
Multipartite entanglement criteria in the vicinity of Dicke states were formulated~\cite{Toth-2007b,Toth-2009,Guhne-2010,Bergmann-2013},
also based on quantum metrology~\cite{Toth-2012,Hyllus-2012,Duan-2011,Vitagliano-2017},
which made possible the experimental detection of multipartite entanglement
in photonic~\cite{Krischek-2011} or cold atomic systems~\cite{Lucke-2011,Hamley-2012,Lucke-2014,Lange-2018}.
Entanglement was also characterized in terms of different entanglement measures
in pure Dicke states and even in the mixtures of symmetric Dicke states
in the qubit~\cite{Stockton-2003,Munizzi-2024}
and also the qu$d$it~\cite{Wei-2003,Popkov-2005,Hayashi-2008,Wei-2008,Zhu-2010} cases.
Dicke states are also important examples in tomography~\cite{Toth-2010c}.
Dicke states of small numbers of qubits
were also prepared directly in quantum optical experiments~\cite{Haffner-2005,Kiesel-2007,Prevedel-2009,Wieczorek-2009},
different methods of preparation were worked out also for quantum computers~\cite{Bartschi-2019,Wang-2021,Bond-2023,Nepomechie-2024a},
and even the matrix product state form of qu$d$it Dicke states could be derived explicitly~\cite{Raveh-2024b}.

Let us recall first some basic notions in \emph{partial separability}~\cite{Dur-1999,Acin-2001,Seevinck-2008,Szalay-2015b}.
A quantum state $\rho$ of a multipartite system $S$ is \emph{$\xi$-separable},
that is, separable with respect to a partition $\xi = \set{X_1,X_2,\dots}$,
if it is a statistical mixture 
of states being product with respect to that partition,
\begin{equation}
    \rho = \sum_i w_i \bigotimes_{X\in \xi} \rho_{X,i},
\end{equation}
where the finite number of weights $w_i$
are nonnegative and sum up to $1$,
$\rho_{X,i}$ are states of subsystems $X\subseteq S$,
which are disjoint and cover the whole system $S$.
We call a state 
\emph{fully separable}, if it is $\xi$-separable and each $X\in\xi$ is an elementary subsystem;
\emph{partition-separable}, if it is $\xi$-separable with respect to a nontrivial partition $\xi$ (containing at least two parts);
and \emph{bipartition-separable}, if it is $\xi$-separable with respect to a bipartition $\xi$ (containing exactly two parts).
It is clear that if a state is partition-separable then it is also bipartition-separable,
since it is separable with respect to all bipartitions $\upsilon =\set{Y,\cmpl{Y}}$ \emph{coarser} than $\xi$,
that is, the subsystems $Y$ and $\cmpl{Y}=S\setminus Y$ are unions of subsystems $X\in\xi$.
This is because we are always allowed not to take into account some of the tensorproduct symbols~\cite{Szalay-2015b},
then the state $\rho$ above can be recast as
\begin{equation}
\label{eq:psbps}
    \sum_i w_i 
        \Bigl(\bigotimes_{\substack{X\in\xi\\ X\subseteq Y}} \rho_{X,i}\Bigr)\otimes 
        \Bigl(\bigotimes_{\substack{X\in\xi\\ X\subseteq \cmpl{Y}}} \rho_{X,i}\Bigr)
    = \sum_i w_i \rho_{Y,i}\otimes \rho_{\cmpl{Y},i}.
\end{equation}
A state is \emph{biseparable}, if it is the mixture of (bi)partition-separable states,
that is, mixture of $\xi$-separable states for \emph{possibly different} nontrivial partitions $\xi$.
\emph{Genuinely multipartite entangled} (GME) states are those which are not biseparable.
In this multipartite case, the \emph{Peres-Horodecki criterion}~\cite{Peres-1996,Horodecki-1996} is about bipartition-separability,
\begin{equation}
\label{eq:Peres}
    \text{$\rho$ is $\set{Y,\cmpl{Y}}$-sep.} \;\;\Longrightarrow\;\; \rho^{\T_Y} \geq0,
\end{equation}
where the \emph{partial transpose} $\T_Y$ is the linear map given on elementary tensors
as $(A_Y\otimes B_{\cmpl{Y}})^{\T_Y} = A_Y^\T\otimes B_{\cmpl{Y}}$,
where $\T$ is the \emph{transpose} of the matrix of the operator $A_Y$ in a fixed basis,
that is, $\bra{i}A_Y^\T\ket{j} = \bra{j}A_Y\ket{i}$.
Although the map $\T_Y$ is given with respect to a fixed local basis,
the positivity of $\rho^{\T_Y}$ and therefore the criterion~\eqref{eq:Peres} are independent of this choice.

We consider \emph{symmetric} state vectors%
\footnote{Symmetric state vectors are state vectors $\ket{\phi}$ for which $P_n\ket{\phi}=\ket{\phi}$,
where $P_n=\frac1{n!}\sum_{\sigma\in\DscGrp{S}_n} R_\sigma$ is the projection operator,
projecting onto the symmetric subspace,
where $R_\sigma$ is the representation of the symmetric group $\DscGrp{S}_n$ on the multipartite Hilbert space,
permuting the elementary subsystems.
The $m$-partite reduced states $\rho_m=\Tr_{n-m}(\proj{\phi})$ 
of an $n$-partite symmetric pure state $\proj{\phi}$ given by the symmetric state vector $\ket{\phi}$
is a \emph{symmetric operator}, that is, it acts only on the symmetric subspace, $P_m\rho_mP_m=\rho_m$.
(Not to be confused with \emph{permutation invariant} operators, defined as $R_\sigma A R_\sigma^\dagger=A$.)
All the possible pure convex decompositions of a symmetric state contain symmetric pure states given by symmetric state vectors.
(Note that the convex decompositions of permutation invariant states is a more difficult problem,
the extremal points are not necessarily pure states.)
See also Refs.~\cite{Harrow-2013,Toth-2009b,Marconi-2025}.}%
, contained in the symmetric subspace of the Hilbert space~\cite{Harrow-2013,Toth-2009b,Marconi-2025}.
The classification of multipartite entanglement simplifies significantly for these states,
namely, these are either fully separable or GME~\cite{GuhneToth-priv,Toth-2009b,Eckert-2002,Ichikawa-2008}.
This is because every pure decomposition of a symmetric state consists of symmetric pure states,
which are known to be either fully separable or fully entangled.
Consequently, a symmetric state entangled with respect to any bipartition is GME,
and a symmetric state separable with respect to any bipartition is fully separable.
In this symmetric multipartite case, the Peres-Horodecki criterion~\eqref{eq:Peres} takes the form
\begin{subequations}
\label{eq:PeresSymm}
    \begin{align}
    \text{symm.~$\rho$ is fully-sep.} \;\;&\Longrightarrow\;\; \forall Y\subset S: \rho^{\T_Y} \geq0,\\
    \intertext{or, contrapositively,}
    \text{symm.~$\rho$ is GME} \;\;&\Longleftarrow\;\; \exists Y\subset S: \rho^{\T_Y} \not\geq0,
\end{align}
\end{subequations}
making possible to identify NPT-GME states directly.
Note that the reverse implications still do not hold,
and symmetric PPT-GME states were also constructed~\cite{Toth-2009b}.

For $n\geq1$, the $n$-qu$d$it symmetric Dicke state vectors
are the equal weight superpositions
of the permutations of the $\ket{i_1,i_2,\dots,i_n}$ elements of a fixed local (tensor product) basis.
(We number the basis vectors from $i=1$ up to $d$ for simplicity.)
It is convenient to label the $n$-qu$d$it symmetric Dicke states with \emph{excitation indices},
also called \emph{occupation numbers},
which are \emph{multiindices} $\mi{n}=(n_1,n_2,\dots,n_d)\in\mathbb{N}^d_0$,
being normalized, $\norm{\mi{n}}:=\sum_{i=1}^d n_i = n$.
Note that $n_i=0$ is also allowed.
Let us have the set of the possible occupation numbers  
\begin{equation}
\label{eq:indexset}
    I^d_n := \bigsset{\mi{n}\in\mathbb{N}^d_0}{\norm{\mi{n}}=n}.
\end{equation}
For such a multiindex $\mi{n}\in I^d_n$,
let us have the (nonnormalized) Dicke vector
\begin{equation}
\label{eq:Dicke_nonnorm}
    \ket{\Tilde{\D}_{\mi{n}}}
    :=\ket{\underbrace{11\dots1}_{n_1}\underbrace{22\dots2}_{n_2}\dots\underbrace{dd\dots d}_{n_d}} +\text{`perms.'},
\end{equation}
where taking all the possible \emph{different} orderings of the basis vectors is understood.
The norm-square of this is the multinomial coefficient 
$\norm{\Tilde{\D}_{\mi{n}}}^2=\scalp{\Tilde{\D}_{\mi{n}}}{\Tilde{\D}_{\mi{n}}}
=\binom{n}{\mi{n}}=\frac{n!}{\prod_{i=1}^d n_i!}$,
leading to the (normalized) \emph{$n$-qu$d$it symmetric Dicke state vector of occupation $\mi{n}\in I^d_n$},
\begin{equation}
\label{eq:Dicke_norm}
    \ket{\D_\mi{n}} := \binom{\norm{\mi{n}}}{\mi{n}}^{-1/2}\ket{\Tilde{\D}_{\mi{n}}}.
\end{equation}
The vectors $\ket{\Tilde{\D}}_{\mi{n}}$ and $\ket{\D_\mi{n}}$ are also called 
elementary symmetric tensors and symmetric basis states~\cite{Hayashi-2008}, respectively,
since $\bigsset{\ket{\D_\mi{n}}}{\mi{n}\in I^d_n}$ is an orthonormal set,
$\scalp{\D_\mi{n}}{\D_{\mi{n}'}} = \delta_{\mi{n},\mi{n}'}$,
and those span the symmetric subspace of the Hilbert space of the $n$-partite composite system~\cite{Harrow-2013}.
The dimension of the symmetric subspace is the $\abs{I^d_n}$ number of the possible occupation numbers,
which is given by the binomial coefficient as
$\abs{I^d_n} = \binom{n+d-1}{d-1} = \binom{n+d-1}{n}=\frac{(n+d-1)!}{n!(d-1)!}$,
coming from elementary combinatorics (see the `stars and bars problem'~\cite{Stanley-2012}).
Note that we also cover the extreme case of one single system, $n=1$,
then it is just the computational basis itself, 
that is, say $n_i=1$ and $n_{j\neq i}=0$ for an $i$,
then $\ket{\D_\mi{n}} = \ket{i}$.
Note also that if there is only one nonzero occupation in a composite system,
say $n_i=n$ and $n_{j\neq i}=0$ for an $i$,
then $\ket{\D_\mi{n}}=\ket{i,i,\dots,i}$, which is fully separable.

For any $1\leq m<n$,
the nonnormalized Dicke vectors~\eqref{eq:Dicke_nonnorm} can be decomposed into two subsystems of sizes $m$ and $n-m$ as
\begin{equation}
\label{eq:Schmidt_nonnorm}
    \ket{\Tilde{\D}_{\mi{n}}}=\sum_{\mi{m}\in I^d_{m,\mi{n}}}\ket{\Tilde{\D}_{\mi{m}}}\otimes\ket{\Tilde{\D}_{\mi{n}-\mi{m}}},
\end{equation}
where the summation runs over the multiindices in the restricted index set
\begin{equation}
\label{eq:indexset_restr}
    I^d_{m,\mi{n}}
    := \bigsset{\mi{m}\in\mathbb{N}_0^d}{\norm{\mi{m}}=m,\mi{m}\leq \mi{n}}
    \subseteq I^d_m,
\end{equation}
where the relation $\leq$ is understood elementwisely.
$I^d_{m,\mi{n}}$ is the intersection of a rectangular hypercuboid specified by $\mi{0}:=(0,0,\dots,0)$ and $\mi{n}$
and the hyperplane $\norm{\mi{m}}=m$,
which is difficult to walk through sequentially if $d\geq3$.
(We provide a proof of the decomposition~\eqref{eq:Schmidt_nonnorm} in the Appendix.
For earlier proofs, see Sec.~3 of Ref.~\cite{Popkov-2005} or Appendix A of Ref.~\cite{Carrasco-2016}.)
Then the \eqref{eq:Schmidt_nonnorm}-like decomposition of the Dicke state vector~\eqref{eq:Dicke_norm} is
\begin{subequations}
\begin{equation}
\label{eq:Schmidt_norm}
     \ket{\D_\mi{n}}
     = \sum_{\mi{m}\in I^d_{m,\mi{n}}} \sqrt{\eta^\mi{n}_\mi{m}}  \ket{\D_\mi{m}} \otimes \ket{\D_{\mi{n}-\mi{m}}},
\end{equation}
where
\begin{equation}
\label{eq:Schmidt_coef}
    \eta^\mi{n}_\mi{m}
    := \frac{\binom{\norm{\mi{m}}}{\mi{m}}\binom{\norm{\mi{n}-\mi{m}}}{\mi{n}-\mi{m}}}{\binom{\norm{\mi{n}}}{\mi{n}}}
    = \eta^\mi{n}_{\mi{n}-\mi{m}}.
\end{equation}
\end{subequations}
Since the Dicke state vectors of the subsystems
$\bigsset{\ket{\D_\mi{m}   }}{\mi{m} \in I^d_{m}  }$ and 
$\bigsset{\ket{\D_{\mi{m}'}}}{\mi{m}'\in I^d_{n-m}}$ form orthonormal bases,
the formula~\eqref{eq:Schmidt_norm} is just the Schmidt decomposition
of the $\ket{\D_\mi{n}}$ Dicke state vector
with the Schmidt coefficients $\eta^\mi{n}_\mi{m}$~\eqref{eq:Schmidt_coef}
and Schmidt rank $\abs{I^d_{m,\mi{n}}}$~\eqref{eq:indexset_restr}.
The reduced states%
\footnote{The reduced states of all subsystems of a given size $m$ are of the same form,
because of the permutation symmetry of the vector~\eqref{eq:Dicke_nonnorm}.
Accordingly, $\Tr_{n-m}$ simply denotes partial trace over any subsystem of size $n-m$.}
of subsystems of size $m$ are then
\begin{equation}
\label{eq:reduced}
    \rho_{\mi{n},m} := \Tr_{n-m}\bigl(\proj{\D_\mi{n}}\bigr)= \sum_{\mi{m}\in I^d_{m,\mi{n}}} \eta^\mi{n}_\mi{m} \proj{\D_\mi{m}}.
\end{equation}
(It also follows that $\sum_{\mi{m}\in I^d_{m,\mi{n}}} \eta^\mi{n}_\mi{m}=1$, being just $\Tr(\rho_{\mi{n},m})$,
which is the multinomial generalization of the Vandermonde identity for binomials~\cite{Stanley-2012}.)
For later use, we would like to cover also the $m=n$ trivial reduction,
then~\eqref{eq:Schmidt_norm} does not make sense,
but~\eqref{eq:reduced} still holds with $I^d_{n,\mi{n}}=\set{\mi{n}}$ and $\eta^\mi{n}_\mi{n}=1$
by the original definitions~\eqref{eq:indexset_restr} and \eqref{eq:Schmidt_coef}, noting that $0!=1$,
and $\rho_{\mi{n},n}=\proj{\D_\mi{n}}$~\eqref{eq:reduced}, as it has to be.

\begin{figure}
    \centering
    \includegraphics[width=.8\columnwidth]{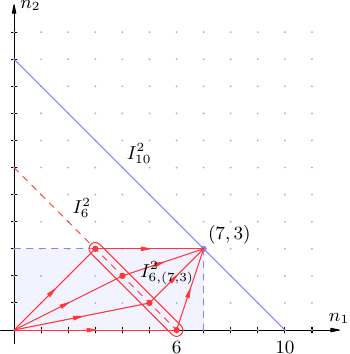}
    \caption{Example of the index sets $I^d_n$~\eqref{eq:indexset} and $I^d_{m,\mi{n}}$~\eqref{eq:indexset_restr}
    for $n=10$ qubits ($d=2$), of occupation $\mi{n}=(7,3)$ for subsystem size $m=6$.
    Arrows illustrate the occupation number multiindices $\mi{m}$ and $\mi{n}-\mi{m}$,
    appearing in the decompositions~\eqref{eq:Schmidt_nonnorm} and~\eqref{eq:Schmidt_norm},
    adding up to $\mi{n}$.}
    \label{fig:I73}
\end{figure}

For example, for qubits ($d=2$), the second component of the $\mi{n}=(n_1,n_2)=:(n-e,e)$ occupation number multiindex,
considered as the number of excitations $\ket{2}$ over the ground state $\ket{1}$,
is usually used to label the states~\cite{Stockton-2003,Guhne-2009}.
Then the multinomial coefficients boil down to the binomial ones, $\binom{n}{\mi{n}}=\binom{n}{e}=\frac{n!}{e!(n-e)!}$,
and 
    $\ket{\D^n_e}:=\ket{\D_{(n-e,e)}}
    =\binom{n}{e}^{-1/2}\bigl(\ket{\underbrace{11\dots1}_{n-e}\underbrace{22\dots2}_{e}}+\text{`perms.'}\bigr)$.
The Schmidt decomposition \eqref{eq:Schmidt_norm} in the qubit case is then
    $\ket{\D^n_e}=\sum^{l_{\text{max}}}_{l=l_{\text{min}}}\sqrt{\binom{m}{l}\binom{n-m}{e-l}/\binom{n}{e}} \ket{\D^m_l}\otimes\ket{\D^{n-m}_{e-l}}$,
where $l$ is the second component of the $\mi{m}=(m-l,l)$ multiindex in~\eqref{eq:Schmidt_nonnorm},
by which one could walk through the index set~\eqref{eq:indexset_restr} sequentially,
$I^2_{m,(n-e,e)} = \bigsset{(m-l,l)}{l_{\text{min}}\leq l \leq l_{\text{max}}}$,
where
$l_{\text{min}} = \text{max}\set{0,e-(n-m)}$ and
$l_{\text{max}} = \text{min}\set{m,e}$.
(For illustrations, see Figure~\ref{fig:I73}.)
As a concrete example, the doubly excited three-qubit Dicke state is 
$\ket{\D^3_2}=\ket{\D_{(1,2)}}=\frac{1}{\sqrt{3}}\bigl(\ket{122}+\ket{212}+\ket{221}\bigr)$,
which is equivalent to the W state~\cite{Dur-2000b}.
We can also illustrate the Schmidt decomposition~\eqref{eq:Schmidt_norm} 
of this for $m=2$ as
$\ket{\D^3_2} = \sqrt{2/3}\ket{\D^2_1}\otimes\ket{\D^1_1}+\sqrt{1/3}\ket{\D^2_2}\otimes\ket{\D^1_0}$,
which is much more expressive with the general multiindex labeling,
$\ket{\D_{(1,2)}}=\sqrt{2/3}\ket{\D_{(1,1)}}\otimes\ket{\D_{(0,1)}}+\sqrt{1/3}\ket{\D_{(0,2)}}\otimes\ket{\D_{(1,0)}}
=\frac{1}{\sqrt{3}}\bigl(\sqrt{2}\frac{1}{\sqrt{2}}(\ket{12}+\ket{21})\otimes\ket{2}+\ket{22}\otimes\ket{1}\bigr)$,
which is indeed $\frac{1}{\sqrt{3}}\bigl(\ket{122}+\ket{212}+\ket{221}\bigr)$.

Let us consider now the $n$-partite Dicke state $\ket{\D_\mi{n}}$ ($n\geq2$)
for an arbitrary occupation number multiindex $\mi{n}\in I^d_n$.
We are interested in the bipartite entanglement inside its $m$-partite subsystems ($2\leq m\leq n$),
with respect to the split into $k$ and $(m-k)$-partite subsystems ($1\leq k\leq m-1$).
We have already seen the rather special property of Dicke states
that their Schmidt vectors are also Dicke states~\eqref{eq:Schmidt_norm}, helping our derivations a lot.
Exploiting this, the reduced state~\eqref{eq:reduced} takes the form
\begin{equation}
\label{eq:reduced2}
\begin{split}
    \rho_{\mi{n},m}
    &= \sum_{\mi{m}\in I^d_{m,\mi{n}}} \eta^\mi{n}_\mi{m}
    \sum_{\mi{k},\mi{k}'\in I^d_{k,\mi{m}}} \sqrt{\eta^\mi{m}_\mi{k}} \sqrt{\eta^\mi{m}_{\mi{k}'}} \\
    &\qquad\qquad\quad\ket{\D_\mi{k}}\bra{\D_{\mi{k}'}} \otimes \ket{\D_{\mi{m}-\mi{k}}}\bra{\D_{\mi{m}-\mi{k}'}},
\end{split}
\end{equation}
convenient for the calculation of the partial transpose.
To detect entanglement~\eqref{eq:PeresSymm}, we have to confirm the nonpositivity of the partial transpose%
\footnote{The partial transposed states in all subsystems of a given size $k$ are of the same form,
because of the permutation symmetry of the vector~\eqref{eq:Dicke_nonnorm}.
Accordingly, $\T_k$ simply denotes partial transpose over any subsystem of size $k$.}
$\rho_{\mi{n},m}^{\T_k}$,
which holds if we find a vector $\ket{\psi}$ giving $\bra{\psi}\rho_{\mi{n},m}^{\T_k}\ket{\psi}<0$.
For the role of $\ket{\psi}$, let us have the educated guess
\begin{equation}
\label{eq:ansatz}
    \ket{\psi} := \alpha \ket{\D_{\hat{\mi{k}}}}  \otimes \ket{\D_{\hat{\mi{m}}-\hat{\mi{k}}'}}
                 + \beta  \ket{\D_{\hat{\mi{k}}'}} \otimes \ket{\D_{\hat{\mi{m}}-\hat{\mi{k}} }}
\end{equation}
with free parameters $\hat{\mi{m}}\in I^d_m$, $\hat{\mi{k}},\hat{\mi{k}}'\in I^d_{k,\hat{\mi{m}}}$ and $\alpha,\beta\in\mathbb{C}$.
Noting that $\bigl(\ket{\D_\mi{k}}\bra{\D_{\mi{k}'}}\bigr)^\T = \ket{\D_{\mi{k}'}}\bra{\D_\mi{k}}$
in the computational basis,
we calculate the sandwich $\bra{\psi}\rho_{\mi{n},m}^{\T_k}\ket{\psi}$ with the state~\eqref{eq:reduced2} and the ansatz~\eqref{eq:ansatz}.
For this first we calculate the sandwich
$\bra{\psi}\bigl(\ket{\D_\mi{k}}\bra{\D_{\mi{k}'}} \otimes \ket{\D_{\mi{m}-\mi{k}'}}\bra{\D_{\mi{m}-\mi{k}}}\bigr)\ket{\psi}$
for the partial transpose of the basis part of the state~\eqref{eq:reduced2},
\begin{widetext} 
\begin{equation*}
\begin{split}
    &\bigl(\overline{\alpha} \bra{\D_{\hat{\mi{k}}}}  \otimes \bra{\D_{\hat{\mi{m}}-\hat{\mi{k}}'}}
         +\overline{\beta}  \bra{\D_{\hat{\mi{k}}'}} \otimes \bra{\D_{\hat{\mi{m}}-\hat{\mi{k}} }}\bigr)
    \Bigl(\ket{\D_\mi{k}}\bra{\D_{\mi{k}'}} \otimes \ket{\D_{\mi{m}-\mi{k}'}}\bra{\D_{\mi{m}-\mi{k}}}\Bigr)
    \bigl( \alpha           \ket{\D_{\hat{\mi{k}}}}  \otimes \ket{\D_{\hat{\mi{m}}-\hat{\mi{k}}'}}
         + \beta            \ket{\D_{\hat{\mi{k}}'}} \otimes \ket{\D_{\hat{\mi{m}}-\hat{\mi{k}} }}\bigr)\\
    &=\overline{\alpha} \alpha 
    \scalp{\D_{\hat{\mi{k}}}}{\D_\mi{k}}
    \scalp{\D_{\mi{k}'}}{\D_{\hat{\mi{k}}}}
    \scalp{\D_{\hat{\mi{m}}-\hat{\mi{k}}'}}{\D_{\mi{m}-\mi{k}'}}
    \scalp{\D_{\mi{m}-\mi{k}}}{\D_{\hat{\mi{m}}-\hat{\mi{k}}'}}
    +\overline{\alpha} \beta
    \scalp{\D_{\hat{\mi{k}}}}{\D_\mi{k}}
    \scalp{\D_{\mi{k}'}}{\D_{\hat{\mi{k}}'}}
    \scalp{\D_{\hat{\mi{m}}-\hat{\mi{k}}'}}{\D_{\mi{m}-\mi{k}'}}
    \scalp{\D_{\mi{m}-\mi{k}}}{\D_{\hat{\mi{m}}-\hat{\mi{k}}}}\\
    &+\overline{\beta} \alpha
    \scalp{\D_{\hat{\mi{k}}'}}{\D_\mi{k}}
    \scalp{\D_{\mi{k}'}}{\D_{\hat{\mi{k}}}}
    \scalp{\D_{\hat{\mi{m}}-\hat{\mi{k}}}}{\D_{\mi{m}-\mi{k}'}}
    \scalp{\D_{\mi{m}-\mi{k}}}{\D_{\hat{\mi{m}}-\hat{\mi{k}}'}}
    +\overline{\beta} \beta
    \scalp{\D_{\hat{\mi{k}}'}}{\D_\mi{k}}
    \scalp{\D_{\mi{k}'}}{\D_{\hat{\mi{k}}'}}
    \scalp{\D_{\hat{\mi{m}}-\hat{\mi{k}}}}{\D_{\mi{m}-\mi{k}'}}
    \scalp{\D_{\mi{m}-\mi{k}}}{\D_{\hat{\mi{m}}-\hat{\mi{k}} }}.
\end{split}
\end{equation*}
To obtain $\bra{\psi}\rho_{\mi{n},m}^{\T_k}\ket{\psi}$,
we have to deal with the summations in the state~\eqref{eq:reduced2}
and the Kronecker-$\delta$s, coming from the orthonormality of the Dicke state vectors above.
(Remember that $\mi{m},\mi{k},\mi{k}'$ are summation indices in the state~\eqref{eq:reduced2},
and $\hat{\mi{m}},\hat{\mi{k}},\hat{\mi{k}}'$ are free-to-choose parameters of the ansatz~\eqref{eq:ansatz}.)\\
For the first term, we have the factors
$\delta_{\hat{\mi{k}},\mi{k}}
\delta_{\mi{k}',\hat{\mi{k}}}
\delta_{\hat{\mi{m}}-\hat{\mi{k}}',\mi{m}-\mi{k}'}
\delta_{\mi{m}-\mi{k},\hat{\mi{m}}-\hat{\mi{k}}'}$,
this is nonzero iff $\mi{k}=\mi{k}'=\hat{\mi{k}}$ and $\mi{m}=\hat{\mi{m}}-\hat{\Delta}$,
where $\hat{\mi{\Delta}}=\hat{\mi{k}}'-\hat{\mi{k}}$.
That is, the parameters $\hat{\mi{m}},\hat{\mi{k}},\hat{\mi{k}}'$ are needed to be such that
$\hat{\mi{m}}-\hat{\mi{\Delta}}\in I^d_{m,\mi{n}}$ and 
$\hat{\mi{k}} \in I^d_{k,\hat{\mi{m}}-\hat{\mi{\Delta}}}$
to meet the indices $\mi{m},\mi{k},\mi{k}'$ in the sum to make this term to be nonzero.\\
For the second term, we have the factors
$\delta_{\hat{\mi{k}},\mi{k}}
\delta_{\mi{k}',\hat{\mi{k}}'}
\delta_{\hat{\mi{m}}-\hat{\mi{k}}',\mi{m}-\mi{k}'}
\delta_{\mi{m}-\mi{k},\hat{\mi{m}}-\hat{\mi{k}}}$,
this is nonzero iff $\mi{k}=\hat{\mi{k}}$, $\mi{k}'=\hat{\mi{k}}'$ and $\mi{m}=\hat{\mi{m}}$.
That is, the parameters $\hat{\mi{m}},\hat{\mi{k}},\hat{\mi{k}}'$ are needed to be such that
$\hat{\mi{m}}\in I^d_{m,\mi{n}}$ and 
$\hat{\mi{k}},\hat{\mi{k}}' \in I^d_{k,\hat{\mi{m}}}$
to meet the indices $\mi{m},\mi{k},\mi{k}'$ in the sum to make this term to be nonzero.\\
The third and the fourth terms can be dealt with similarly,
and finally we have
\begin{equation*}
\begin{split}
    &\bra{\psi}\rho_{\mi{n},m}^{\T_k}\ket{\psi} 
    = \sum_{\mi{m}\in I^d_{m,\mi{n}}} \eta^\mi{n}_\mi{m}
    \sum_{\mi{k},\mi{k}'\in I^d_{k,\mi{m}}} \sqrt{\eta^\mi{m}_\mi{k}} \sqrt{\eta^\mi{m}_{\mi{k}'}}
    \bra{\psi}
    \Bigl(\ket{\D_\mi{k}}\bra{\D_{\mi{k}'}} \otimes \ket{\D_{\mi{m}-\mi{k}'}}\bra{\D_{\mi{m}-\mi{k}}}\Bigr) 
    \ket{\psi}\\
    &=\overline{\alpha} \alpha 
    \eta_{\hat{\mi{m}}-\hat{\mi{\Delta}}}^{\mi{n}} \eta^{\hat{\mi{m}}-\hat{\mi{\Delta}}}_{\hat{\mi{k}}} \delta\bigl(\hat{\mi{m}}-\hat{\mi{\Delta}}\in I^d_{m,\mi{n}}\bigr) 
    \delta\bigl(\hat{\mi{k}} \in I^d_{k,\hat{\mi{m}}-\hat{\mi{\Delta}}}\bigr)
    +\overline{\alpha} \beta 
    \eta^{\mi{n}}_{\hat{\mi{m}}} \sqrt{\eta^{\hat{\mi{m}}}_{\hat{\mi{k}}} \eta^{\hat{\mi{m}}}_{\hat{\mi{k}}'}}
    \delta\bigl(\hat{\mi{m}}\in I^d_{m,\mi{n}}\bigr)
    \delta\bigl(\hat{\mi{k}}\in I^d_{k,\hat{\mi{m}}}\bigr) 
    \delta\bigl(\hat{\mi{k}}'\in I^d_{k,\hat{\mi{m}}}\bigr) \\
    &+\overline{\beta} \alpha 
    \eta^{\mi{n}}_{\hat{\mi{m}}} \sqrt{\eta^{\hat{\mi{m}}}_{\hat{\mi{k}}} \eta^{\hat{\mi{m}}}_{\hat{\mi{k}}'}} 
    \delta\bigl(\hat{\mi{m}}\in I^d_{m,\mi{n}}\bigr)
    \delta\bigl(\hat{\mi{k}}\in I^d_{k,\hat{\mi{m}}}\bigr)
    \delta\bigl(\hat{\mi{k}}'\in I^d_{k,\hat{\mi{m}}}\bigr)
    +\overline{\beta} \beta 
    \eta_{\hat{\mi{m}}+\hat{\mi{\Delta}}}^{\mi{n}} \eta^{\hat{\mi{m}}+\hat{\mi{\Delta}}}_{\hat{\mi{k}}'} \delta\bigl(\hat{\mi{m}}+\hat{\mi{\Delta}}\in I^d_{m,\mi{n}}\bigr) 
    \delta\bigl(\hat{\mi{k}}' \in I^d_{k,\hat{\mi{m}}+\hat{\mi{\Delta}}}\bigr),
\end{split}
\end{equation*}
where the symbol $\delta(\mi{a} \in A)$ gives $1$ if $\mi{a} \in A$ and $0$ otherwise. 
This expression is a Hermitian form of nonnegative coefficients in the two complex variables $\alpha$ and $\beta$,
which can take negative values if and only if its discriminant 
\begin{equation*}
\begin{split}
    \eta_{\hat{\mi{m}}-\hat{\mi{\Delta}}}^{\mi{n}}
    \eta_{\hat{\mi{m}}+\hat{\mi{\Delta}}}^{\mi{n}}
    \eta^{\hat{\mi{m}}-\hat{\mi{\Delta}}}_{\hat{\mi{k}}}
    \eta^{\hat{\mi{m}}+\hat{\mi{\Delta}}}_{\hat{\mi{k}}'} 
    \delta\bigl(\hat{\mi{m}}-\hat{\mi{\Delta}}\in I^d_{m,\mi{n}}\bigr)
    &\delta\bigl(\hat{\mi{m}}+\hat{\mi{\Delta}}\in I^d_{m,\mi{n}}\bigr) 
    \delta\bigl(\hat{\mi{k}} \in I^d_{k,\hat{\mi{m}}-\hat{\mi{\Delta}}}\bigr) 
    \delta\bigl(\hat{\mi{k}}' \in I^d_{k,\hat{\mi{m}}+\hat{\mi{\Delta}}}\bigr)  \\ 
    &\qquad\qquad- (\eta^{\mi{n}}_{\hat{\mi{m}}})^2
    \eta^{\hat{\mi{m}}}_{\hat{\mi{k}}}
    \eta^{\hat{\mi{m}}}_{\hat{\mi{k}}'}\delta\bigl(\hat{\mi{m}}\in I^d_{m,\mi{n}}\bigr)
    \delta\bigl(\hat{\mi{k}}\in I^d_{k,\hat{\mi{m}}}\bigr)
    \delta\bigl(\hat{\mi{k}}'\in I^d_{k,\hat{\mi{m}}}\bigr)
\end{split}
\end{equation*}
is negative.
For this, the second term is necessarily nonvanishing,
so the parameters in \eqref{eq:ansatz} are restricted to 
${\hat{\mi{m}}\in I^d_{m,\mi{n}}}$ and 
$\hat{\mi{k}},\hat{\mi{k}}'\in I^d_{k,\hat{\mi{m}}}$.
It is easy to see that in this case
$\hat{\mi{k}} \in I^d_{k,\hat{\mi{m}}-\hat{\mi{\Delta}}}$ and
$\hat{\mi{k}}'\in I^d_{k,\hat{\mi{m}}+\hat{\mi{\Delta}}}$ also hold
(e.g., if $\hat{\mi{k}}'\leq\hat{\mi{m}}$ then $\mi{0}\leq\hat{\mi{m}}-\hat{\mi{k}}'$,
so $\hat{\mi{k}}\leq\hat{\mi{m}}-\hat{\mi{k}}'+\hat{\mi{k}}$),
so the negativity condition on the discriminant reads as
\begin{equation*}
    \eta_{\hat{\mi{m}}-\hat{\mi{\Delta}}}^{\mi{n}}
    \eta_{\hat{\mi{m}}+\hat{\mi{\Delta}}}^{\mi{n}} 
    \eta^{\hat{\mi{m}}-\hat{\mi{\Delta}}}_{\hat{\mi{k}}}
    \eta^{\hat{\mi{m}}+\hat{\mi{\Delta}}}_{\hat{\mi{k}}'}\\
    \delta\bigl(\hat{\mi{m}}-\hat{\mi{\Delta}}\in I^d_{m,\mi{n}}\bigr)
    \delta\bigl(\hat{\mi{m}}+\hat{\mi{\Delta}}\in I^d_{m,\mi{n}}\bigr)
    < (\eta^{\mi{n}}_{\hat{\mi{m}}})^2 \eta^{\hat{\mi{m}}}_{\hat{\mi{k}}} \eta^{\hat{\mi{m}}}_{\hat{\mi{k}}'},
\end{equation*}
assuming ${\hat{\mi{m}}\in I^d_{m,\mi{n}}}$ and 
$\hat{\mi{k}},\hat{\mi{k}}'\in I^d_{k,\hat{\mi{m}}}$.
Substituting the Schmidt coefficients~\eqref{eq:Schmidt_coef}, the inequality is simplified,
\begin{equation*}
    \binom{n-m}{\mi{n}-\hat{\mi{m}}+\hat{\mi{\Delta}}}
    \binom{n-m}{\mi{n}-\hat{\mi{m}}-\hat{\mi{\Delta}}}
    \delta\bigl(\hat{\mi{m}}-\hat{\mi{\Delta}}\in I^d_{m,\mi{n}}\bigr)
    \delta\bigl(\hat{\mi{m}}+\hat{\mi{\Delta}}\in I^d_{m,\mi{n}}\bigr)
    < \binom{n-m}{\mi{n}-\hat{\mi{m}}}^2.
\end{equation*}
This obviously holds if any of the two conditions $\hat{\mi{m}}\pm\hat{\mi{\Delta}}\in I^d_{m,\mi{n}}$ is violated,
however, in some cases this may not be guaranteed,
so we proceed in a different way.
If there are at least two nonzero occupations in $\mi{n}$, say $n_i,n_j\neq 0$,
then we can always choose
$\mi{m}\in I^d_{m,\mi{n}}$ such that $m_i,m_j\neq 0$,
and then
$\hat{\mi{k}},\hat{\mi{k}}'\in I^d_{k,\hat{\mi{m}}}$ such that they differ only in those two positions,
so $\hat{\Delta}_i=1$, $\hat{\Delta}_j=-1$ and $0$ elsewhere,
for which the inequality reads as
\begin{equation*}
    \frac{(n-m)!}{(n_i-\hat{m}_i+1)!(n_j-\hat{m}_j-1)!}
    \frac{(n-m)!}{(n_i-\hat{m}_i-1)!(n_j-\hat{m}_j+1)!}
    < \biggl( \frac{(n-m)!}{(n_i-\hat{m}_i)!(n_j-\hat{m}_j)! }\biggr)^2,
\end{equation*}
which is
\begin{equation*}
    \frac{n_i-\hat{m}_i}{n_i-\hat{m}_i+1} \frac{n_j-\hat{m}_j}{n_j-\hat{m}_j+1} <1,
\end{equation*}
which holds for every $\hat{\mi{m}}\in I^d_{m,\mi{n}}$.
This concludes the proof.
\end{widetext} 

Summing up, we have that 
if there are at least two nonzero occupations in $\mi{n}$,
then the $m$-partite reduced Dicke state $\rho_{\mi{n},m} = \Tr_{n-m}\bigl(\proj{\D_\mi{n}}\bigr)$ for all $2\leq m \leq n$
is NPT for all splits, $\rho_{\mi{n},m}^{\T_k}\not\geq0$ for all $1\leq k\leq m-1$,
then the Peres-Horodecki criterion for symmetric states~\eqref{eq:PeresSymm} tells us that it is NPT-GME.
(Note that this holds also for the original Dicke state $\proj{\D_\mi{n}}$, which is the $m=n$ case,
although this is easier to see directly by the Schmidt rank $\abs{I^d_{k,\mi{n}}}>1$, see~\eqref{eq:indexset_restr}.)
Otherwise, if there is only one occupation in $\mi{n}$, then the Dicke state
and also all the reduced Dicke states are fully separable pure states $\proj{i,i,\dots,i}$.

Note that for some small systems,
mixtures of Dicke states are known to be entangled \emph{if and only if} NPT,
so there is no PPT-entanglement,
however, this does not hold in general~\cite{Yu-2016,Tura-2018,Romero-Palleja-2025}.
PPT-entanglement is considered to be a weaker form of entanglement,
based on that
PPT-entangled states are bound-entangled, that is, no pure entangled states can be distilled from them~\cite{Horodecki-1998},
and it is conjectured that their Schmidt number, although scaling linearly with the local dimension~\cite{Huber-2018,Pal-2019,Krebs-2024}, cannot be maximal.
(For qutrits, this conjecture~\cite{Sanpera-2001} has been proven~\cite{Yang-2016}.)
Although we have seen that the reduced Dicke states are NPT-entangled,
there are further important open questions about the strength of this entanglement
compared to the weakness of PPT-entanglement,
e.g., regarding Schmidt number or distillability.

\textit{Acknowledgment:} We thank Otfried Gühne, Géza Tóth and Iagoba Apellaniz for discussions.
Financial support 
of the \emph{Hungarian National Research, Development and Innovation Office}
within the grant K-134983,
within the `Frontline' Research Excellence Programme KKP-133827,
and within the Quantum Information National Laboratory of Hungary;
and of the \emph{Wigner Research Centre for Physics}
within the Wigner Internship program are gratefully acknowledged.
Sz.~Sz.~happily acknowledges the support of the wonderful Bach performances of Marta Czech and Sir András Schiff.

\textit{Appendix:} To see the decomposition~\eqref{eq:Schmidt_nonnorm},
we have by construction that
(i) for any $\mi{m}\in I^d_{m,\mi{n}}$, the vector $\ket{\Tilde{\D}_{\mi{m}}}\otimes\ket{\Tilde{\D}_{\mi{n}-\mi{m}}}$
is the linear combination of basis vectors $\ket{i_1,i_2,\dots,i_d}$ with coefficients $+1$ (there are no repetitions);
(ii) for any different $\mi{m},\mi{m}'\in I^d_{m,\mi{n}}$, the vectors 
$\ket{\Tilde{\D}_{\mi{m}}}\otimes\ket{\Tilde{\D}_{\mi{n}-\mi{m}}}$ and
$\ket{\Tilde{\D}_{\mi{m}'}}\otimes\ket{\Tilde{\D}_{\mi{n}-\mi{m}'}}$ contain different basis vectors;
(iii) every basis vector in $\ket{\Tilde{\D}_{\mi{n}}}$ is contained in a $\ket{\Tilde{\D}_{\mi{m}}}\otimes\ket{\Tilde{\D}_{\mi{n}-\mi{m}}}$ for an $\mi{m}\in I^d_{m,\mi{n}}$;
(iv) every basis vector in every $\ket{\Tilde{\D}_{\mi{m}}}\otimes\ket{\Tilde{\D}_{\mi{n}-\mi{m}}}$ is contained in $\ket{\Tilde{\D}_{\mi{n}}}$.

\bibliographystyle{quantum}
\bibliography{Dicke}{}

\begin{thebibliography}{10}

\bibitem{Werner-1989}
Reinhard~F. Werner.
\newblock ``Quantum states with {E}instein-{P}odolsky-{R}osen correlations
  admitting a hidden-variable model''.
\newblock \href{https://dx.doi.org/10.1103/PhysRevA.40.4277}{Phys. Rev. A {\bf
  40}, 4277}~(1989).

\bibitem{Horodecki-2009}
Ryszard Horodecki, Pawe\l{} Horodecki, Micha\l{} Horodecki, and Karol
  Horodecki.
\newblock ``Quantum entanglement''.
\newblock \href{https://dx.doi.org/10.1103/RevModPhys.81.865}{Rev. Mod. Phys.
  {\bf 81}, 865}~(2009).

\bibitem{Schrodinger-1935a}
Ervin Schr{\"o}dinger.
\newblock ``{D}ie gegenwärtige {S}ituation in der {Q}uantenmechanik''.
\newblock \href{https://dx.doi.org/10.1007/BF01491891}{Naturwissenschaften {\bf
  23}, 807}~(1935).

\bibitem{Guhne-2009}
Otfried G\"{u}hne and Géza Tóth.
\newblock ``Entanglement detection''.
\newblock \href{https://dx.doi.org/10.1016/j.physrep.2009.02.004}{Phys. Rep.
  {\bf 474}, 1}~(2009).

\bibitem{Szalay-2011}
{\relax Sz}il{\'a}rd {\relax Sz}alay.
\newblock ``Separability criteria for mixed three-qubit states''.
\newblock \href{https://dx.doi.org/10.1103/PhysRevA.83.062337}{Phys. Rev. A
  {\bf 83}, 062337}~(2011).

\bibitem{Peres-1996}
Asher Peres.
\newblock ``Separability criterion for density matrices''.
\newblock \href{https://dx.doi.org/10.1103/PhysRevLett.77.1413}{Phys. Rev.
  Lett. {\bf 77}, 1413}~(1996).

\bibitem{Horodecki-1996}
Michał Horodecki, Paweł Horodecki, and Ryszard Horodecki.
\newblock ``Separability of mixed states: necessary and sufficient
  conditions''.
\newblock \href{https://dx.doi.org/10.1016/S0375-9601(96)00706-2}{Phys. Lett. A
  {\bf 223}, 1}~(1996).

\bibitem{Horodecki-1997}
Pawel Horodecki.
\newblock ``Separability criterion and inseparable mixed states with positive
  partial transposition''.
\newblock \href{https://dx.doi.org/10.1016/S0375-9601(97)00416-7}{Phys. Lett. A
  {\bf 232}, 333}~(1997).

\bibitem{Huber-2018}
Marcus Huber, Ludovico Lami, C\'ecilia Lancien, and Alexander M\"uller-Hermes.
\newblock ``High-dimensional entanglement in states with positive partial
  transposition''.
\newblock \href{https://dx.doi.org/10.1103/PhysRevLett.121.200503}{Phys. Rev.
  Lett. {\bf 121}, 200503}~(2018).

\bibitem{Pal-2019}
K\'aroly~F. P\'al and Tam\'as V\'ertesi.
\newblock ``Class of genuinely high-dimensionally-entangled states with a
  positive partial transpose''.
\newblock \href{https://dx.doi.org/10.1103/PhysRevA.100.012310}{Phys. Rev. A
  {\bf 100}, 012310}~(2019).

\bibitem{Krebs-2024}
Robin Krebs and Mariami Gachechiladze.
\newblock ``High {Schmidt} number concentration in quantum bound entangled
  states''.
\newblock \href{https://dx.doi.org/10.1103/PhysRevLett.132.220203}{Phys. Rev.
  Lett. {\bf 132}, 220203}~(2024).

\bibitem{Bengtsson-2006}
Ingemar Bengtsson and Karol {\r Z}yczkowski.
\newblock ``Geometry of quantum states: An introduction to quantum
  entanglement''.
\newblock \href{https://dx.doi.org/10.1017/CBO9780511535048}{Cambridge
  University Press}. ~(2006).

\bibitem{Bengtsson-2017b}
Ingemar Bengtsson and Karol {\r Z}yczkowski.
\newblock ``Geometry of quantum states: An introduction to quantum
  entanglement''.
\newblock \href{https://dx.doi.org/10.1017/9781139207010}{Chapter 17, A brief
  introduction to multipartite entanglement}.
\newblock Cambridge University Press. ~(2017).
\newblock 2nd edition.
\newblock  \href{http://arxiv.org/abs/1612.07747}{arXiv:1612.07747}.

\bibitem{Horodecki-2024}
Pawel Horodecki, {\L}ukasz Rudnicki, and Karol {\r Z}yczkowski.
\newblock ``Encyclopedia of mathematical physics''.
\newblock \href{https://dx.doi.org/10.48550/arXiv.2409.04566}{Chapter
  Multipartite entanglement}.
\newblock Elsevier. ~(2025).
\newblock 2nd edition.
\newblock  \href{http://arxiv.org/abs/2409.04566}{arXiv:2409.04566}.

\bibitem{Dur-1999}
Wolfgang Dür, J.~Ignacio Cirac, and Rolf Tarrach.
\newblock ``Separability and distillability of multiparticle quantum systems''.
\newblock \href{https://dx.doi.org/10.1103/PhysRevLett.83.3562}{Phys. Rev.
  Lett. {\bf 83}, 3562}~(1999).

\bibitem{Acin-2001}
Antonio Ac\'{i}n, Dagmar Bru\ss{}, Maciej Lewenstein, and Anna Sanpera.
\newblock ``Classification of mixed three-qubit states''.
\newblock \href{https://dx.doi.org/10.1103/PhysRevLett.87.040401}{Phys. Rev.
  Lett. {\bf 87}, 040401}~(2001).

\bibitem{Seevinck-2008}
Michael Seevinck and Jos Uffink.
\newblock ``Partial separability and entanglement criteria for multiqubit
  quantum states''.
\newblock \href{https://dx.doi.org/10.1103/PhysRevA.78.032101}{Phys. Rev. A
  {\bf 78}, 032101}~(2008).

\bibitem{Szalay-2015b}
{\relax Sz}il{\'a}rd {\relax Sz}alay.
\newblock ``Multipartite entanglement measures''.
\newblock \href{https://dx.doi.org/10.1103/PhysRevA.92.042329}{Phys. Rev. A
  {\bf 92}, 042329}~(2015).

\bibitem{Szalay-2019}
{\relax Sz}ilárd {\relax Sz}alay.
\newblock ``$k$-stretchability of entanglement, and the duality of
  $k$-separability and $k$-producibility''.
\newblock \href{https://dx.doi.org/10.22331/q-2019-12-02-204}{{Quantum} {\bf
  3}, 204}~(2019).

\bibitem{Szalay-2025}
{\relax Sz}ilárd {\relax Sz}alay and Géza Tóth.
\newblock ``Alternatives of entanglement depth and metrological entanglement
  criteria''.
\newblock \href{https://dx.doi.org/10.22331/q-2025-04-18-1718}{{Quantum} {\bf
  9}, 1718}~(2025).

\bibitem{Dur-2000b}
Wolfgang Dür, Guifre Vidal, and J.~Ignacio Cirac.
\newblock ``Three qubits can be entangled in two inequivalent ways''.
\newblock \href{https://dx.doi.org/10.1103/PhysRevA.62.062314}{Phys. Rev. A
  {\bf 62}, 062314}~(2000).

\bibitem{Dicke-1954}
R.~H. Dicke.
\newblock ``Coherence in spontaneous radiation processes''.
\newblock \href{https://dx.doi.org/10.1103/PhysRev.93.99}{Phys. Rev. {\bf 93},
  99}~(1954).

\bibitem{Toth-2007b}
Géza Tóth.
\newblock ``Detection of multipartite entanglement in the vicinity of symmetric
  {Dicke} states''.
\newblock \href{https://dx.doi.org/10.1364/JOSAB.24.000275}{J. Opt. Soc. Am. B
  {\bf 24}, 275}~(2007).

\bibitem{Toth-2009}
Géza Tóth, Witlef Wieczorek, Roland Krischek, Nikolai Kiesel, Patrick
  Michelberger, and Harald Weinfurter.
\newblock ``Practical methods for witnessing genuine multi-qubit entanglement
  in the vicinity of symmetric states''.
\newblock \href{https://dx.doi.org/10.1088/1367-2630/11/8/083002}{New J. Phys.
  {\bf 11}, 083002}~(2009).

\bibitem{Guhne-2010}
Otfried Gühne and Michael Seevinck.
\newblock ``Separability criteria for genuine multiparticle entanglement''.
\newblock \href{https://dx.doi.org/10.1088/1367-2630/12/5/053002}{New J. Phys.
  {\bf 12}, 053002}~(2010).

\bibitem{Bergmann-2013}
Marcel Bergmann and Otfried Gühne.
\newblock ``Entanglement criteria for {Dicke} states''.
\newblock \href{https://dx.doi.org/10.1088/1751-8113/46/38/385304}{J. Phys. A:
  Math. Theor. {\bf 46}, 385304}~(2013).

\bibitem{Toth-2012}
Géza Tóth.
\newblock ``Multipartite entanglement and high-precision metrology''.
\newblock \href{https://dx.doi.org/10.1103/PhysRevA.85.022322}{Phys. Rev. A
  {\bf 85}, 022322}~(2012).

\bibitem{Hyllus-2012}
Philipp Hyllus, Wies\l{}aw Laskowski, Roland Krischek, Christian Schwemmer,
  Witlef Wieczorek, Harald Weinfurter, Luca Pezzé, and Augusto Smerzi.
\newblock ``Fisher information and multiparticle entanglement''.
\newblock \href{https://dx.doi.org/10.1103/PhysRevA.85.022321}{Phys. Rev. A
  {\bf 85}, 022321}~(2012).

\bibitem{Duan-2011}
Lu-Ming Duan.
\newblock ``Entanglement detection in the vicinity of arbitrary {Dicke}
  states''.
\newblock \href{https://dx.doi.org/10.1103/PhysRevLett.107.180502}{Phys. Rev.
  Lett. {\bf 107}, 180502}~(2011).

\bibitem{Vitagliano-2017}
Giuseppe Vitagliano, Iagoba Apellaniz, Matthias Kleinmann, Bernd Lücke,
  Carsten Klempt, and Géza Tóth.
\newblock ``Entanglement and extreme spin squeezing of unpolarized states''.
\newblock \href{https://dx.doi.org/10.1088/1367-2630/19/1/013027}{New J. Phys.
  {\bf 19}, 013027}~(2017).

\bibitem{Krischek-2011}
Roland Krischek, Christian Schwemmer, Witlef Wieczorek, Harald Weinfurter,
  Philipp Hyllus, Luca Pezzé, and Augusto Smerzi.
\newblock ``Useful multiparticle entanglement and sub-shot-noise sensitivity in
  experimental phase estimation''.
\newblock \href{https://dx.doi.org/10.1103/PhysRevLett.107.080504}{Phys. Rev.
  Lett. {\bf 107}, 080504}~(2011).

\bibitem{Lucke-2011}
Bernd Lücke, Manuel Scherer, Jens Kruse, Luca Pezzé, Frank Deuretzbacher,
  Phillip Hyllus, Oliver Topic, Jan Peise, Wolfgang Ertmer, Jan Arlt, Luis
  Santos, Augusto Smerzi, and Carsten Klempt.
\newblock ``Twin matter waves for interferometry beyond the classical limit''.
\newblock \href{https://dx.doi.org/10.1126/science.1208798}{Science {\bf 334},
  773}~(2011).

\bibitem{Hamley-2012}
Chris~D. Hamley, Corey~S. Gerving, Thai~M. Hoang, Eva~M. Bookjans, and
  Michael~S. Chapman.
\newblock ``Spin-nematic squeezed vacuum in a quantum gas''.
\newblock \href{https://dx.doi.org/10.1038/nphys2245}{Nature Physics {\bf 8},
  305}~(2012).

\bibitem{Lucke-2014}
Bernd Lücke, Jan Peise, Giuseppe Vitagliano, Jan Arlt, Luis Santos, Géza
  Tóth, and Carsten Klempt.
\newblock ``Detecting multiparticle entanglement of {Dicke} states''.
\newblock \href{https://dx.doi.org/10.1103/PhysRevLett.112.155304}{Phys. Rev.
  Lett. {\bf 112}, 155304}~(2014).

\bibitem{Lange-2018}
Karsten Lange, Jan Peise, Bernd Lücke, Ilka Kruse, Giuseppe Vitagliano, Iagoba
  Apellaniz, Matthias Kleinmann, Géza Tóth, and Carsten Klempt.
\newblock ``Entanglement between two spatially separated atomic modes''.
\newblock \href{https://dx.doi.org/10.1126/science.aao2035}{Science {\bf 360},
  416}~(2018).

\bibitem{Stockton-2003}
John~K. Stockton, J.~M. Geremia, Andrew~C. Doherty, and Hideo Mabuchi.
\newblock ``Characterizing the entanglement of symmetric many-particle
  spin-$\frac{1}{2}$ systems''.
\newblock \href{https://dx.doi.org/10.1103/PhysRevA.67.022112}{Phys. Rev. A
  {\bf 67}, 022112}~(2003).

\bibitem{Munizzi-2024}
William Munizzi and Howard~J. Schnitzer.
\newblock ``Entropy cones and entanglement evolution for {Dicke } states''.
\newblock \href{https://dx.doi.org/10.1103/PhysRevA.109.012405}{Phys. Rev. A
  {\bf 109}, 012405}~(2024).

\bibitem{Wei-2003}
Tzu-Chieh Wei and Paul~M. Goldbart.
\newblock ``Geometric measure of entanglement and applications to bipartite and
  multipartite quantum states''.
\newblock \href{https://dx.doi.org/10.1103/PhysRevA.68.042307}{Phys. Rev. A
  {\bf 68}, 042307}~(2003).

\bibitem{Popkov-2005}
Vladislav Popkov, Mario Salerno, and Gunter Schütz.
\newblock ``Entangling power of permutation-invariant quantum states''.
\newblock \href{https://dx.doi.org/10.1103/PhysRevA.72.032327}{Phys. Rev. A
  {\bf 72}, 032327}~(2005).

\bibitem{Hayashi-2008}
Masahito Hayashi, Damian Markham, Mio Murao, Masaki Owari, and Shashank
  Virmani.
\newblock ``Entanglement of multiparty-stabilizer, symmetric, and antisymmetric
  states''.
\newblock \href{https://dx.doi.org/10.1103/PhysRevA.77.012104}{Phys. Rev. A
  {\bf 77}, 012104}~(2008).

\bibitem{Wei-2008}
Tzu-Chieh Wei.
\newblock ``Relative entropy of entanglement for multipartite mixed states:
  Permutation-invariant states and dür states''.
\newblock \href{https://dx.doi.org/10.1103/PhysRevA.78.012327}{Phys. Rev. A
  {\bf 78}, 012327}~(2008).

\bibitem{Zhu-2010}
Huangjun Zhu, Lin Chen, and Masahito Hayashi.
\newblock ``Additivity and non-additivity of multipartite entanglement
  measures''.
\newblock \href{https://dx.doi.org/10.1088/1367-2630/12/8/083002}{New J. Phys.
  {\bf 12}, 083002}~(2010).

\bibitem{Toth-2010c}
Géza Tóth, Witlef Wieczorek, David Gross, Roland Krischek, Christian
  Schwemmer, and Harald Weinfurter.
\newblock ``Permutationally invariant quantum tomography''.
\newblock \href{https://dx.doi.org/10.1103/PhysRevLett.105.250403}{Phys. Rev.
  Lett. {\bf 105}, 250403}~(2010).

\bibitem{Haffner-2005}
Hartmut H\"{a}ffner, Wolfgang H\"{a}nsel, Christian~F. Roos, Jan Benhelm, Dany
  Chek-al kar, Michael Chwalla, Timo K\"{o}rber, Umakant~D. Rapol, Mark Riebe,
  Piet~O. Schmidt, Christoph Becher, Otfried G\"{u}hne, Wolfgang D\"{u}r, and
  Rainer Blatt.
\newblock ``Scalable multiparticle entanglement of trapped ions''.
\newblock \href{https://dx.doi.org/10.1038/nature04279}{Nature {\bf 438},
  643}~(2005).

\bibitem{Kiesel-2007}
Nikolai Kiesel, Christian Schmid, Géza Tóth, Enrique Solano, and Harald
  Weinfurter.
\newblock ``Experimental observation of four-photon entangled {Dicke} state
  with high fidelity''.
\newblock \href{https://dx.doi.org/10.1103/PhysRevLett.98.063604}{Phys. Rev.
  Lett. {\bf 98}, 063604}~(2007).

\bibitem{Prevedel-2009}
Robert Prevedel, Gunther Cronenberg, Mark~S. Tame, Mauro Paternostro, Philip
  Walther, Myungshik Kim, and Anton Zeilinger.
\newblock ``Experimental realization of {Dicke} states of up to six qubits for
  multiparty quantum networking''.
\newblock \href{https://dx.doi.org/10.1103/PhysRevLett.103.020503}{Phys. Rev.
  Lett. {\bf 103}, 020503}~(2009).

\bibitem{Wieczorek-2009}
Witlef Wieczorek, Roland Krischek, Nikolai Kiesel, Patrick Michelberger, Géza
  Tóth, and Harald Weinfurter.
\newblock ``Experimental entanglement of a six-photon symmetric {Dicke}
  state''.
\newblock \href{https://dx.doi.org/10.1103/PhysRevLett.103.020504}{Phys. Rev.
  Lett. {\bf 103}, 020504}~(2009).

\bibitem{Bartschi-2019}
Andreas B{\"a}rtschi and Stephan Eidenbenz.
\newblock ``Deterministic preparation of {Dicke} states''.
\newblock In Leszek~Antoni G{\k{a}}sieniec, Jesper Jansson, and Christos
  Levcopoulos, editors, Fundamentals of Computation Theory.
\newblock \href{https://dx.doi.org/10.1007/978-3-030-25027-0_9}{Page 126}.
\newblock Springer, Cham~(2019).

\bibitem{Wang-2021}
Yang Wang and Barbara~M. Terhal.
\newblock ``Preparing {Dicke} states in a spin ensemble using phase
  estimation''.
\newblock \href{https://dx.doi.org/10.1103/PhysRevA.104.032407}{Phys. Rev. A
  {\bf 104}, 032407}~(2021).

\bibitem{Bond-2023}
Liam~J. Bond, Matthew~J. Davis, Ji\ifmmode \check{r}\else~\v{r}\fi{}\'{\i}
  Min\'a\ifmmode~\check{r}\else \v{r}\fi{}, Rene Gerritsma, Gavin~K. Brennen,
  and Arghavan Safavi-Naini.
\newblock ``Global variational quantum circuits for arbitrary symmetric state
  preparation''.
\newblock \href{https://dx.doi.org/10.1103/PhysRevResearch.7.L022072}{Phys.
  Rev. Res. {\bf 7}, L022072}~(2025).

\bibitem{Nepomechie-2024a}
Rafael~I. Nepomechie and David Raveh.
\newblock ``Qudit {Dicke} state preparation''.
\newblock \href{https://dx.doi.org/10.26421/qic24.1-2-2}{Quantum Inf. Comput.
  {\bf 24}, 37}~(2024).

\bibitem{Raveh-2024b}
David Raveh and Rafael~I. Nepomechie.
\newblock ``Dicke states as matrix product states''.
\newblock \href{https://dx.doi.org/10.1103/PhysRevA.110.052438}{Phys. Rev. A
  {\bf 110}, 052438}~(2024).

\bibitem{Harrow-2013}
Aram~W. Harrow.
\newblock ``The church of the symmetric subspace''~(2013)
  \href{http://arxiv.org/abs/1308.6595}{arXiv:1308.6595}.

\bibitem{Toth-2009b}
Géza Tóth and Otfried Gühne.
\newblock ``Entanglement and permutational symmetry''.
\newblock \href{https://dx.doi.org/10.1103/PhysRevLett.102.170503}{Phys. Rev.
  Lett. {\bf 102}, 170503}~(2009).

\bibitem{Marconi-2025}
Carlo Marconi, Guillem M\"{u}ller-Rigat, Jordi Romero-Pallejà, Jordi Tura, and
  Anna Sanpera.
\newblock ``Symmetric quantum states: a review of recent progress''~(2025)
  \href{http://arxiv.org/abs/2506.10185}{arXiv:2506.10185}.

\bibitem{GuhneToth-priv}
Otfried Gühne and Géza Tóth.
\newblock ``private communication''~(2025).

\bibitem{Eckert-2002}
Kai Eckert, John Schliemann, Dagmar Bruß, and Maciej Lewenstein.
\newblock ``Quantum correlations in systems of indistinguishable particles''.
\newblock \href{https://dx.doi.org/10.1006/aphy.2002.6268}{Annals of Physics
  {\bf 299}, 88}~(2002).

\bibitem{Ichikawa-2008}
Tsubasa Ichikawa, Toshihiko Sasaki, Izumi Tsutsui, and Nobuhiro Yonezawa.
\newblock ``Exchange symmetry and multipartite entanglement''.
\newblock \href{https://dx.doi.org/10.1103/PhysRevA.78.052105}{Phys. Rev. A
  {\bf 78}, 052105}~(2008).

\bibitem{Stanley-2012}
Richard~P. Stanley.
\newblock ``Enumerative combinatorics, volume 1''.
\newblock Volume~49 of Cambridge studies in advanced mathematics.
\newblock Cambridge University Press. ~(2012).
\newblock 2nd edition.

\bibitem{Carrasco-2016}
José~A. Carrasco, Federico Finkel, Artemio González-López, Miguel~A.
  Rodríguez, and Piergiulio Tempesta.
\newblock ``Generalized isotropic {Lipkin–Meshkov–Glick} models: ground
  state entanglement and quantum entropies''.
\newblock \href{https://dx.doi.org/10.1088/1742-5468/2016/03/033114}{J. Stat.
  Mech.: Theor. Exp. {\bf 2016}, 033114}~(2016).

\bibitem{Yu-2016}
Nengkun Yu.
\newblock ``Separability of a mixture of {Dicke} states''.
\newblock \href{https://dx.doi.org/10.1103/PhysRevA.94.060101}{Phys. Rev. A
  {\bf 94}, 060101}~(2016).

\bibitem{Tura-2018}
Jordi Tura, Albert Aloy, Ruben Quesada, Maciej Lewenstein, and Anna Sanpera.
\newblock ``Separability of diagonal symmetric states: a quadratic conic
  optimization problem''.
\newblock \href{https://dx.doi.org/10.22331/q-2018-01-12-45}{{Quantum} {\bf 2},
  45}~(2018).

\bibitem{Romero-Palleja-2025}
Jordi Romero-Pallejà, Jennifer Ahiable, Alessandro Romancino, Carlo Marconi,
  and Anna Sanpera.
\newblock ``Multipartite entanglement in the diagonal symmetric subspace''.
\newblock \href{https://dx.doi.org/10.1063/5.0240964}{J. Math. Phys. {\bf 66},
  022203}~(2025).

\bibitem{Horodecki-1998}
Micha\l{} Horodecki, Pawe\l{} Horodecki, and Ryszard Horodecki.
\newblock ``Mixed-state entanglement and distillation: Is there a ``bound''
  entanglement in nature?''.
\newblock \href{https://dx.doi.org/10.1103/PhysRevLett.80.5239}{Phys. Rev.
  Lett. {\bf 80}, 5239}~(1998).

\bibitem{Sanpera-2001}
Anna Sanpera, Dagmar Bru\ss{}, and Maciej Lewenstein.
\newblock ``Schmidt-number witnesses and bound entanglement''.
\newblock \href{https://dx.doi.org/10.1103/PhysRevA.63.050301}{Phys. Rev. A
  {\bf 63}, 050301}~(2001).

\bibitem{Yang-2016}
Yu~Yang, Denny~H. Leung, and Wai-Shing Tang.
\newblock ``All $2$-positive linear maps from {$M_3(\mathbb{C})$} to
  {$M_3(\mathbb{C})$} are decomposable''.
\newblock
  \href{https://dx.doi.org/https://doi.org/10.1016/j.laa.2016.03.050}{Lin. Alg.
  Appl. {\bf 503}, 233}~(2016).

\end{thebibliography}

\end{document}